\documentclass[11pt]{article}
\usepackage{graphicx}
\usepackage{epsfig}

\newcommand{\BABARPubYear}    {00}

\newcommand{\BABARProcNumber} {19}
\newcommand{\SLACPubNumber} {8697}

\def\lbabar{\mbox{{\large\sl B}\hspace{-0.4em} {\normalsize\sl A}\hspace{-0.03em}{\large\sl B}\hspace{-0.4em} {\normalsize\sl A\hspace{-0.02em}R}}}
\usepackage{relsize}
\def\babar{\mbox{\slshape B\kern-0.1em{\smaller A}\kern-0.1em
    B\kern-0.1em{\smaller A\kern-0.2em R}}}

\def\ellp       {\ensuremath{\ell^+}}

\def\Kbar  {\kern 0.2em\overline{\kern -0.2em K}{}}

\def\Kzb   {\ensuremath{\Kbar^0}}
\def\KzKzb {\ensuremath{K^0 \kern -0.16em \Kzb}}
\def\Dz    {\ensuremath{D^0}}
\def\Dbar  {\kern 0.2em\overline{\kern -0.2em D}{}}

\def\Dzb   {\ensuremath{\Dbar^0}}
\def\DzDzb {\ensuremath{D^0 {\kern -0.16em \Dzb}}}

\def\Bz    {\ensuremath{B^0}}

\def\Bbar  {\kern 0.18em\overline{\kern -0.18em B}{}}

\def\Bzb   {\ensuremath{\Bbar^0}}

\def\BzBzb {\ensuremath{B^0 {\kern -0.16em \Bzb}}}

\def\jpsi  {\ensuremath{{J\mskip -3mu/\mskip -2mu\psi\mskip 2mu}}} 
\mathchardef\Upsilon="7107
\def\Y#1S{\ensuremath{\Upsilon{(#1S)}}}

\def\FourS {\Y4S}

\mathchardef\Deltares="7101
\mathchardef\Xi="7104
\mathchardef\Lambda="7103
\mathchardef\Sigma="7106
\mathchardef\Omega="710A
\def\Deltabar   {\kern 0.25em\overline{\kern -0.25em \Deltares}{}}
\def\Lbar {\kern 0.2em\overline{\kern -0.2em\Lambda\kern 0.05em}\kern-0.05em{}}
\def\Sigbar{\kern 0.2em\overline{\kern -0.2em \Sigma}{}}
\def\Xibar{\kern 0.2em\overline{\kern -0.2em \Xi}{}}
\def\Obar{\kern 0.2em\overline{\kern -0.2em \Omega}{}}
\def\Nbar{\kern 0.2em\overline{\kern -0.2em N}{}}
\def\Xbar{\kern 0.2em\overline{\kern -0.2em X}{}}

\def\upsbb {\ensuremath{\Upsilon{\rm( 4S)}\to B\Bbar}}

\def\ev   {\ensuremath{\rm \,e\kern -0.08em V}}
\def\kev  {\ensuremath{\rm \,ke\kern -0.08em V}} 
\def\mev  {\ensuremath{\rm \,Me\kern -0.08em V}} 
\def\gev  {\ensuremath{\rm \,Ge\kern -0.08em V}} 
\def\gevc {\ensuremath{{\rm \,Ge\kern -0.08em V\!/}c}} 
\def\tev  {\ensuremath{\rm \,Te\kern -0.08em V}}
\def\mevc {\ensuremath{{\rm \,Me\kern -0.08em V\!/}c}} 
\def\gevcc{\ensuremath{{\rm \,Ge\kern -0.08em V\!/}c^2}} 
\def\mevcc{\ensuremath{{\rm \,Me\kern -0.08em V\!/}c^2}}

\def\mum  {\ensuremath{\,\mu\rm m}} 

\def\invfb   {\ensuremath{\mbox{\,fb}^{-1}}}
\def\mus  {\ensuremath{\rm \,\mus}}

\def\ps   {\ensuremath{\rm \,ps}}

\def\mus        {\ensuremath{\,\mu{\rm s}}}    
\def\ps         {\ensuremath{{\rm \,ps}}}   
\def\gsim{{~\raise.15em\hbox{$>$}\kern-.85em
          \lower.35em\hbox{$\sim$}~}}
\def\lsim{{~\raise.15em\hbox{$<$}\kern-.85em
          \lower.35em\hbox{$\sim$}~}}

\def\to                 {\ensuremath{\rightarrow}}

\def\pep2{PEP-II}
\def\BF{$B$ Factory}

\providecommand{\eqref}[1]{Eq.~(\ref{eq:#1})}

\def\jetset74   {\mbox{\tt Jetset \hspace{-0.5em}7.\hspace{-0.2em}4}}

\def\dm {\ensuremath{\Delta m_{B^0}}}

\def\delz {\ensuremath{\Delta{ z}}}

\def\dt {\ensuremath{\Delta t}}

\def\re_eb {\ensuremath{Re(\varepsilon_B)}}
\def\gevsq  {\ensuremath{\mbox{\,Ge\kern -0.08em V}^2}} 

\def\BDstarpi{$\Bz\rightarrow D^{*-} \pi^{+}$}
\def\BDstarlnu{$\Bz\rightarrow D^{*-} \ell^+ \nu_{\ell}$ }
\def\DsptoDz{\mbox{$D^{*-}\rightarrow D^{0} \pi^-$}}
\newcommand{\mnusq} {\ifmmode{{M_\nu}^2} \else {$M_{\nu}^2$}\fi} 

\long\def\inst#1{\par\nobreak\kern 4pt\nobreak
    {\it #1}\par\vskip 10pt plus 3pt minus 3pt}

\begin{document}
{\pagestyle{empty}

\begin{flushright}
SLAC-PUB-\SLACPubNumber \\
\babar-PROC-\BABARPubYear/\BABARProcNumber \\
August, 2000 \\
\end{flushright}

\par\vskip 4cm

\begin{center}
\Large \bf 
 Preliminary \babar\ results on \Bz\ mixing with dileptons 
and on lifetime with partially reconstructed \Bz\ decays
\end{center}
\bigskip

\begin{center}
\large 
Christophe Y\`eche\\
(for the \lbabar\ Collaboration)\\
CEA Saclay, DAPNIA/SPP, Bat 141, 91191 Gif-Sur-Yvette cedex, 
France\\
E-mail: yeche@slac.stanford.edu\\
\end{center}
\bigskip \bigskip

\begin{center}
\large \bf Abstract
\end{center}
With an integrated  luminosity
 of  7.7 \invfb\ collected on resonance by  \babar\ at the PEP-II asymmetric \BF,  
we measure the difference in mass between the neutral $B$ eigenstates, 
$\dm$, to be $(0.507\pm 0.015\pm 0.022)\times 10^{12}\,\hbar\, s^{-1}$ with dileptons events 
and present preliminary results for the \Bz\ lifetime,  
$\tau_{B^0} = 1.55 \pm 0.05 \pm 0.07\,\ps$ and
 $ \tau_{B^0} = 1.62 \pm 0.02 \pm 0.09\,\ps$ obtained  from partial reconstruction of 
 the two $B^0$ decay processes respectively \mbox{$B^0\rightarrow D^{*-}\pi^+$} and 
\mbox{$B^0\rightarrow D^{*-} \ell^+ \nu_{\ell}$}.

\vfill
\begin{center}
Contributed to the Proceedings of the 30$^{th}$ International 
Conference on High Energy Physics, \\
7/27/2000---8/2/2000, Osaka, Japan
\end{center}

\vspace{1.0cm}
\begin{center}
{\em Stanford Linear Accelerator Center, Stanford University, 
Stanford, CA 94309} \\ \vspace{0.1cm}\hrule\vspace{0.1cm}
Work supported in part by Department of Energy contract DE-AC03-76SF00515.
\end{center}

\setlength\columnsep{0.20truein}
\twocolumn
\def\sloppy{\tolerance=100000\hfuzz=\maxdimen\vfuzz=\maxdimen}
\sloppy
\vbadness=12000
\hbadness=12000
\flushbottom
\def\figurebox#1#2#3{%
  	\def\arg{#3}%
  	\ifx\arg\empty
  	{\hfill\vbox{\hsize#2\hrule\hbox to #2{\vrule\hfill\vbox to #1{\hsize#2\vfill}\vrule}\hrule}\hfill}%
  	\else
   	{\hfill\epsfbox{#3}\hfill}%
  	\fi}
%

\section{Introduction}
This paper presents three analyses designed to select large samples of \Bz\ mesons using
 inclusive reconstruction techniques. The first method proposes a precise measurement 
of the mixing parameter
$\dm$ using direct dilepton events which represent  4\%  of the $\upsbb$ decays. 
The two other methods provide a measurement of the \Bz\ lifetime
by selecting  \BDstarpi\ and \BDstarlnu\ decays; while the two techniques are different 
in detail, they both share the common feature of making no attempt to reconstruct the $D^0$ 
produced in the  \DsptoDz\ decay, thereby achieving high efficiency compared 
to the exclusive reconstruction.

\section{Measurement of \Bz\ mixing with dileptons}
\subsection{Selection of dilepton events and determination of $\Delta t$ }
\label{dilep_sel}
In this study\cite{BabarPub0010} of the oscillation frequency $\dm$, 
the flavor of the $B$ meson at decay is 
determined by the sign of leptons produced in semileptonic $B$ decays. For this analysis, 
electron and muon candidates are required to pass the {\em{very tight}} selection criteria 
fully described in reference\cite{BabarPub0017}; electrons are essentially selected by specific 
requirements on  energy deposited in the Electromagnetic Calorimeter and muons are identified 
by the use of information provided by the the Instrumented Flux Return.\\
Non $B\bar B$ events (radiative Bhabhas, two-photon and continuum events) are suppressed by
applying cuts on  the Fox-Wolfram ratio of second to zeroth order moments, on the event squared 
invariant mass,  the event aplanarity  and the number of charged tracks. Finally, events with a
lepton coming from \jpsi\ decays  are rejected.\\
The discrimination between direct and cascade leptons  is based on a neural network
which combines five discriminating variables, all calculated in the \FourS\ 
center of mass system:  the  momenta of the two leptons with highest momenta,
the total visible energy, the missing momentum of the event and the opening angle between 
the two leptons.\\ 
The combined effect of the above cuts gives, from Monte Carlo events, a signal purity of 78\%.
The  main source of background consists  of $B\bar B$ events (12\% direct-cascade events). 
The total number of selected on-resonance events is 36631 (10742 electron pairs, 7836 muon 
pairs, and 18053 electron-muon pairs).

\begin{figure*}[hbt]
\begin{center}
\mbox{\epsfig{file=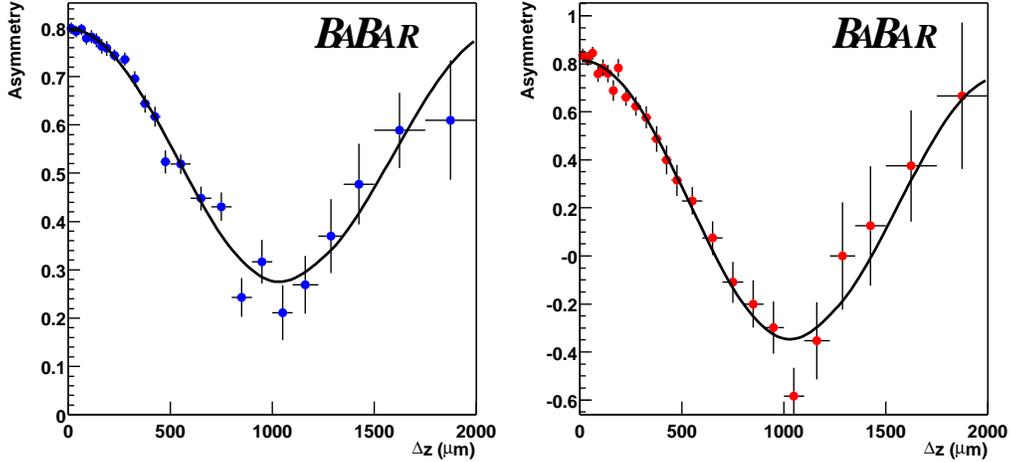,height=6cm}}
\end{center}
\caption{ Time-dependent asymmetry $A_{\ell\ell}(|\dt|)$ for (a) the inclusive dilepton sample
and (b) the dilepton sample enriched with soft pions with a method similar of 
section~\ref{sec:partial}. 
The curve represents the result of the fit.}
\label{asy}
\end{figure*}

\noindent
The $z$ coordinate of the $B$ decay vertex  is determined by taking the $z$ position of the 
point of closest approach of the track to an estimate of the position for $\Upsilon (4{\rm S})$ decay,
obtained by minimizing a $\chi^2$ based on the relative position of the tracks and the beam
spot in tranverse plane. A two-Gaussian fit to the resulting $\delz$ resolution 
function from simulated dilepton events gives $\sigma_n = 87$\mum\ and $\sigma_w = 195$\mum\ 
for the narrow and wide Gaussian, respectively, with 76\% of the events in the narrow Gaussian. 
Then, the time difference between the two $B$ decay times is defined as 
$\dt = \delz/( <\beta\gamma> c)$, with $<\beta\gamma>=0.554$.

\subsection{Measurement of \dm}

The value of $\dm$ is extracted with a $\chi^2$ minimization fit to the dilepton asymmetry:
\begin{equation}
A_{\ell\ell}(|\dt|)=\frac{N(\ell^+,\ell^-) - N(\ell^\pm,\ell^\pm)}
{N(\ell^+,\ell^-) + N(\ell^\pm,\ell^\pm)}.
\label{basicAsy}
\end{equation} 
The fit function  takes into account the various time distributions of the dilepton signal and 
the cascade lepton and the non-$B\overline{B}$ background. The time dependence of this last
and its absolute normalization is obtained from off-resonance data. In the fit, three additional
parameters are left free: the fraction of charged $B$, the mistag fraction and the 
time-dependence of the mistagged events. From a data sample equivalent to $7.73\, fb^{-1}$,
we obtain $\dm=(0.507\pm 0.015\pm 0.022)\times 10^{12}\,\hbar\, s^{-1}$ (see  Figure~\ref{asy})
, the main sources of 
systematic uncertainties are related to the time dependence of the cascade and mis-identified
lepton and to the uncertainty on the resolution function (see  Table~\ref{sys_table}).

\begin{table} [htb]
{\small
\begin{center}
\caption{  Systematic uncertainty on \dm}
\begin{tabular}{|l|c|} 
\hline
Source of & $\sigma (\dm )$   \\
systematic uncertainty & $(10^{12}\,\hbar\, s^{-1})$  \\
\hline
\hline
Non-$B\overline{B}$ background  & 0.005 \\ 
\hline
Mis-Identification  & 0.011 \\ 
\hline
Cascade events  & 0.009 \\ 
\hline
Boost approximation & 0.001  \\
\hline
 Beam spot motion ($\le 20$\mum)  & 0.001 \\ 
\hline
\delz\ resolution function & 0.009  \\
\hline
Tails of the resolution function & 0.004  \\
\hline
Time-dependence of the  & \\
resolution function & 0.006  \\
\hline
Sensitivity to $\Gamma^+$ and $\Gamma^0$ & \\
 (PDG 98 $\pm 1\sigma$) & 0.010   \\
\hline
\hline
Total  & 0.022  \\ 
\hline
\end{tabular}
\label{sys_table} 
\end{center}
}
\end{table}

\vspace*{-0.8cm}
\section{Measurement of  \Bz\ lifetime  with partially reconstructed \Bz}
\label{sec:partial}

\subsection{Event selection and determination of $\Delta z$ }

In  studies\cite{BabarPub0009} of the decays \BDstarpi\ and \BDstarlnu\ reported here,
no attempt is made to reconstruct the \Dz\ decays. Thus, in the hadronic channel, a search 
 is made for a pair of oppositely-charged pions ($\pi_f$, $\pi_s$) and, 
assuming that their origin is a \Bz\ meson and using the beam energy as a constraint, 
calculates the missing  mass $M_{miss}$. This should be the \Dz\ mass if
the hypothesis was correct. The signal region is taken to be the interval 
$M_{miss} > 1.854\,\gevcc$. In the case of the semileptonic decay, due to the limited 
phase space 
available in the decay \DsptoDz,  the $D^{*-}$ four-momentum can be computed by 
approximating its polar and azimuth angles with those of the slow pion $\pi_s$, and 
parametrizing its momentum as a linear function of the $\pi_s$ momentum. Then a cut is 
applied on the invariant mass of the neutrino \mnusq\ estimated  from the \Bz, $D^{*-}$ 
and $\ellp$ four-momenta ($\mnusq > -2\,(\gevcc)^2$).\\
The methods for rejection of the non $B\bar B$ background and the identification of the lepton
are very similar to those described in  section~\ref{dilep_sel}. For the
\BDstarpi, the combinatorial background is reduced by using a Fisher discriminant 
method combining topological variables relating the position of the tracks and 
the pseudo-direction of  the $D^0$.\\ 
The $z$ of the first \Bz\ is obtained by fitting a vertex between  the slow pion and 
the fast pion or the direct lepton with the beam spot constrained. A fit with the
other tracks outside an exclusion cone around the $D^0$ is performed to determine
the $z$ of the second \Bz.

\subsection{$\tau_{\Bz}$ measurement}
The \Bz\ lifetime is determined by means of an unbinned maximum likelihood fit, 
accounting for the event-by-event error determined by the vertex reconstruction 
algorithm. The fit function is the sum of the probability density function (pdf) 
for \Bz\, $B^{\pm}$, and combinatorial background. 
The different features of this background 
(time-dependence, resolution function, etc) are deduced both from Monte Carlo and data with
 wrong charge association. The  results  are:\\
\begin{center}
$\tau_{\Bz} = 1.55 \pm 0.05 \pm 0.07 \ps \quad (D^*\pi),$ \\
$\tau_{\Bz} = 1.62 \pm 0.02 \pm 0.09 \ps \quad (D^*\ell\nu).$\\
\end{center}
The dominant systematics come from the uncertainty on the fraction and the time-dependence
of the backgrounds, the resolution function and the bias due to tracks related to the
unconstructed $D^0$.

\begin{figure}[hbt]
\begin{center}
\mbox{\epsfig{file=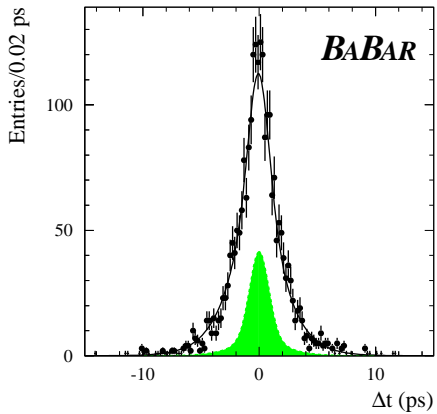,height=6.cm}}
\end{center}
\vspace*{-0.8cm}
\caption{Distribution of $\Delta t$ in ps for ($D^*\pi$) events.
The continuous line shows the result of the fit  and the shaded area 
shows the contribution of the combinatorial background (27\% of the sample). 
} 
\label{lifetime_pi}
\end{figure}

\vspace*{-0.8cm}
\section{Conclusions}
We  present a preliminary study of the  \BzBzb\ oscillation frequency with an inclusive 
sample of dilepton events where an accuracy  already comparable with the 
current world average is obtained. The partial reconstuction of $D^{*-}\pi^+$ and 
$D^{*-} \ell^+ \nu_{\ell}$ gives measurements of the \Bz\ lifetime and 
will allow another determination of \dm\ in the future.

\end{document}